\documentclass[a4paper,fleqn,usenatbib,useAMS]{mnras}

\usepackage{color}

\usepackage{graphicx}	
\usepackage{amsmath}	
\usepackage{amssymb}	
\usepackage{multicol}        
\usepackage{bm}		
\usepackage{pdflscape}	

\usepackage{xcolor}

\defcitealias{santos-lima_etal_2014}{SL+14}
\defcitealias{gary_etal_2000}{G+00}

\usepackage[T1]{fontenc}
\usepackage{ae,aecompl}

\title[Limits on the ions temperature anisotropy in turbulent 
intracluster medium]{Limits on the ions temperature anisotropy in turbulent \\
intracluster medium}

\author[R. Santos-Lima et al.]{
	R. Santos-Lima,$^{1,2,3}$\thanks{Contact e-mail: \href{mailto:reinaldo.santos.de.lima@desy.de}{reinaldo.santos.de.lima@desy.de}}
	H. Yan,$^{1,2}$\thanks{Contact e-mail: \href{mailto:huirong.yan@desy.de}{huirong.yan@desy.de}}
	E. M. de Gouveia Dal Pino,$^{3}$	and A. Lazarian$^{4}$\\
$^{1}$DESY, Platanenallee 6, 15738 Zeuthen, Germany \\
$^{2}$Institut fur Physik und Astronomie, Universit\"{a}t Potsdam, 14476 Potsdam-Golm, Germany \\
$^{3}$Instituto de Astronomia, Geof\'isica e Ci\^encias Atmosf\'ericas, 
Universidade de S\~ao Paulo, R. do Mat\~ao, 1226, S\~ao Paulo, SP 05508-090, Brazil \\
$^{4}$Department of Astronomy, University of Wisconsin, 475 North Charter Street, Madison, WI 53706, USA}

\date{Last updated 2015 May 22; in original form 2013 September 5}

\pubyear{2016}

\begin{document}
\label{firstpage}
\pagerange{\pageref{firstpage}--\pageref{lastpage}}
\maketitle

\begin{abstract}
Turbulence in the weakly collisional intracluster medium of galaxies (ICM) is able to generate
strong thermal velocity anisotropies in the ions (with respect to the local magnetic field direction), if the
magnetic moment of the particles is conserved in the absence of Coulomb collisions. In
this scenario, the anisotropic pressure magnetohydrodynamic (AMHD) turbulence shows a very different statistical behaviour from the 
standard MHD one and is unable to amplify seed magnetic fields, in disagreement with
previous cosmological MHD simulations which are successful to explain the observed magnetic fields 
in the ICM. 
On the other hand, temperature anisotropies can also drive plasma instabilities
which can relax the anisotropy. 
This work aims to compare the relaxation rate with the growth rate of the anisotropies driven by the turbulence. 
We employ quasilinear
theory to estimate the ions scattering rate 
due to the  parallel firehose, mirror, and
ion-cyclotron instabilities, for a set of 
plasma parameters resulting from AMHD
simulations of the turbulent ICM. We show that the ICM turbulence can sustain only anisotropy levels
very close to the instabilities thresholds. 
We argue that the AMHD model which bounds the anisotropies at the marginal stability levels 
can describe the Alfvenic turbulence cascade in the ICM. 

\end{abstract}




\begingroup
\let\clearpage\relax
\endgroup
\newpage

\graphicspath{{./figs/}}

\section{Introduction}
\label{sec:introduction}

The intracluster medium of galaxies (ICM) is composed by a plasma weakly collisional and magnetized, 
with turbulent motions at the large scales. The cosmological mergers of sub-clusters are thought to be the major 
sources of turbulence in the ICM.
The turbulence time-scale for the ICM is 
$\tau_{turb} \sim 10^{16} \; {\rm sec}$ 
(using $L_{turb} = 500 \; {\rm kpc}$ and $U_{turb} = 10^{3} \; {\rm km/s}$
as the length scale and velocity of the largest scale turbulent motions)
while the time-scale for the Coulomb collisions between ions is estimated as
$\tau_i \sim 10^{15} \; {\rm sec}$~\footnote{Considering
the ions temperature $T_{i} = 10 \; {\rm keV}$, 
density $n = 10^{-3} \; {\rm cm}^{-3}$, and the Coulomb logarithm $\ln \Lambda = 20$.}
(mean-free-path of $\sim30 \; {\rm kpc}$ for the ion-ion collisions), which requires
a nearly collisionless approach.

The conservation of the first adiabatic invariant of the charged particles
(magnetic moment) combined with
the large scale plasma motions 
stretching/compressing the magnetic fields in the ICM leads to the development of differences 
between the parallel (to the local field lines) and the gyro component of the thermal velocities of the ions.
Therefore, 
being highly turbulent and weakly collisional, 
the ICM naturally develops anisotropies in the local distribution of the ions thermal velocities.
This anisotropy is a source of free energy, which can
 trigger electromagnetic plasma instabilities 
(such as ion-cyclotron, mirror, and firehose;
see for example \citealt{gary_1993})
playing a very important role on the dynamics of the system 
and on the turbulence evolution itself \citep{schekochihin_cowley_2006, kowal_etal_2011, 
santos-lima_etal_2014, mogavero_schekochihin_2014}.
Nonetheless, such a role remains still poorly understood. 

An anisotropic magnetohydrodynamic (AMHD) approximation that assumes a bi-Maxwellian distribution of thermal velocities, i.e., takes into account two independent temperature components 
(one for the thermal velocity parallel to the local magnetic field and another for the gyro-motion of the particles), 
can be employed in this situation. 
The solutions of the AMHD equations reveals some linear instabilities
(mirror and firehose), corresponding to the large wave-length (fluid) limit of these plasma instabilities
(see for example \citealt{hau_wang_2007}; \citealt{kowal_etal_2011}).

The effects of anisotropy driven instabilities at the micro-scales are still a matter of debate. 
In one scenario, the plasma instabilities 
saturate the anisotropy at low levels, close to the instabilities thresholds 
(see e.g. \citealt{mogavero_schekochihin_2014}).
In another scenario, if
the anisotropy survives during the dynamical time-scales 
and anisotropic thermal stresses dominate the dynamics of the system, 
there is a change in the traditional MHD turbulence picture with the presence of instabilities 
at fluid scales. 
Studies of the turbulence statistics and the magnetic field amplification applying the last scenario 
to galaxy clusters \citep{kowal_etal_2011, santos-lima_etal_2014, falceta-goncalves_kowal_2015} 
as well as to the Earth's magnetosphere 
\citep{meng_etal_2012} revealed drastic differences compared to the isotropic MHD approach.
Nevertheless in this case the numerical description is incomplete, as the instabilities 
that should develop at the subgrid scales may influence the large scale anisotropy evolution 
(see \citealt{mogavero_schekochihin_2014} and discussion in Section~\ref{sec:applicability_ba}). 

All these collisionless effects possibly influence the Cosmic-Ray (CR) propagation and acceleration in the 
ICM. For instance, compressible modes rather than Alfv\'{e}nic turbulence have been identified as the dominant 
agent for particle acceleration \citep{yan_lazarian_2002}. In the absence of the anomalous scattering 
of the ions produced by the kinetic instabilities, the large parallel viscosity of the ions 
will damp efficiently the compressible modes in the ICM. At the same time, if magnetic 
fluctuations caused by the temperature anisotropy are present in the large scale ICM, 
they could have direct impact on the propagation of Cosmic-Rays in the medium 
(e.g. \citealt{nakwacki_peralta-ramos_2013}).

Obviously, the effects of the plasma instabilities at the kinetic scales
cannot be captured by any MHD model
(see discussion in Section~\ref{sec:applicability_ba} about the  
general effects of the subgrid phenomena).
The impact of a fast thermal relaxation due to particle scattering by the kinetic instabilities on the turbulence cascade 
and on the magnetic field amplification was also investigated in \citet{santos-lima_etal_2014}, 
where the rate of this process was considered as a free-parameter. 
The pitch-angle scattering rate caused by some of these instabilities has been investigated 
for the solar wind via two-dimensional PIC and hybrid (PIC-MHD) simulations \citep{gary_etal_2000} 
and also via a quasilinear approach \citep{seough_yoon_2012, yoon_seough_2012},
and the results point to a scattering time of the order of the linear growth rate of the instabilities 
(which can be $\sim$ ion kinetic times-scales).
In fact, these studies only considered the evolution of the instabilities starting from  
an unstable anisotropy level (see Section~\ref{sec:limitations_ql}).
In the situation of a very slow 
driving of the thermal anisotropy (compared to the ion thermal gyrofrequency),
recent two-dimensional PIC simulations \citep{riquelme_etal_2012, riquelme_etal_2015} and
hybrid \citep{kunz_etal_2014, melville_etal_2016} have demonstrated that the 
anisotropy relaxation arising from the instabilities do not necessarily result
in instantaneous anomalous scattering of the ions 
during the time of anisotropy driving by the turbulent motions
(see Section~\ref{sec:slow_driving}).

Nonetheless, a self-consistent treatment of the feedback of these instabilities connected to the turbulence cascade is still missing. 
A guiding procedure was developed relating consistently both plasma instabilities 
induced by high energy CR (gyroresonance instability) 
and the turbulence in the interstellar and intergalactic media \citep{lazarian_beresnyak_2006, yan_lazarian_2011}.

The aim of this work is to evaluate the limits on the temperature anisotropy particularly in 
the turbulent intergalactic or intra-cluster medium taking into account the scattering produced by the 
electromagnetic instabilities 
triggered by temperature anisotropy in an approach similar to the work by \citet{yan_lazarian_2011}. 
For this goal, we will compare directly 
the ions scattering rate obtained from quasilinear theory with the 
anisotropy generation rate by turbulence obtained from AMHD simulations \citep{santos-lima_etal_2014}.

This study is organized as follows: in \S\ref{sec:empirical_bounds} we review the observed 
relation between the bounds on the temperature anisotropy in the solar wind 
and the collisionless instabilities;
in \S\ref{sec:amhd_simul} we describe briefly the AMHD simulations
used in this work, and in \S\ref{sec:ql} we present the quasilinear equations employed for calculating 
the scattering rate of the ions; in \S\ref{sec:results} we present the results.  
In \S\ref{sec:discussion} we discuss some limitations and consequences of our study and we 
relate it to other works; and finally in \S\ref{sec:summary} we summarize and conclude 
our analysis.

\section{Empirical bounds on the temperature anisotropy}
\label{sec:empirical_bounds}

The distribution function of the thermal velocities of the species in the nearly collisionless plasmas 
of the 
Earth's magnetosphere is accessible via direct measurements by spacecrafts.
The data accumulated from the last decades have shown that the electrons and ions in the solar wind at a distance $\approx 1\; {\rm AU}$ 
present a bi-Maxwellian distribution, with 
the maximum anisotropy in the temperatures 
anti-correlated with the local plasma $\beta$, which is the ratio between the thermal 
and magnetic energy densities (see more details in \citealt{marsch_2006} and references therein; 
\citealt{hellinger_etal_2006, stverak_etal_2008}).
These limits on the anisotropy degree 
are below the expected levels when one assumes
adiabatic conservation of the magnetic momentum of the particles $p_{\perp}/B$ 
(where $p_{\perp}$ is here the perpendicular momentum of the particle and $B$ is the intensity of the magnetic field)
during the expansion/compression of the solar wind (see for example \citealt{bale_etal_2009}).

These limits are interpreted as resulting from the non-linear saturation of the
kinetic instabilities driven by the temperature anisotropy \citep{gary_1993}.
The linear dispersion of a plasma with one or more species having a bi-Maxwellian distribution 
presents a few instabilities resulting from the temperature anisotropy. 
The observed limits on the temperature anisotropy have been identified with the 
approximate thresholds for the 
firehose, mirror, and ion-cyclotron instabilities (see for example \citealt{hellinger_etal_2006, 
bale_etal_2009, maruca_etal_2012}).

The physical process limiting the temperature anisotropy 
depends on the specific instability and on the initial anisotropy level (see discussion in Sections~\ref{sec:limitations_ql} and~\ref{sec:slow_driving}). 
After the instabilities growth saturation, this process is understood in terms of collisionless dissipation, 
with particles being scattered by the collective electromagnetic fluctuations caused by the instabilities \citep{kunz_etal_2014}.
These wave-particle interactions (quasi-collisions) diffuse the momentum of the particles an so their pitch angle, 
relaxing the distribution function towards a Maxwellian one.
This effect is not only observed in the solar wind, but also in laboratory plasmas \citep{keiter_1999} 
and in fully non-linear plasma simulations 
\citep{tajima_etal_1977, tanaka_1993, gary_etal_1997, gary_etal_1998, gary_etal_2000, le_etal_2010, nishimura_etal_2002,  
riquelme_etal_2012, kunz_etal_2014, riquelme_etal_2015, sironi_narayan_2015, sironi_2015}.

\section{Temperature anisotropy development in the turbulent ICM: AMHD simulations}
\label{sec:amhd_simul}

In \citet[SL+14 hereafter]{santos-lima_etal_2014}, a numerical study of the ICM turbulence was 
carried out by means of 
anisotropic MHD simulations of forced turbulence in a periodic box. The temperature 
anisotropy evolution was modeled via the CGL closure \citep{chew_etal_1956} 
modified by the addition of a phenomenological pitch-angle scattering term:
\begin{equation}
	\frac{\partial A}{\partial t} = \left( \frac{\partial A}{\partial t} \right)_{CGL} + \left( \frac{\partial A}{\partial t} \right)_{scatt},
\end{equation}
\begin{equation}
	\label{eq:dadt_cgl}
	\left( \frac{\partial A}{\partial t} \right)_{CGL} =
	- \nabla \cdot \left( A \bmath{u} \right) 
	+ 3 A \bmath{b} \cdot \left[ \left( \bmath{b} \cdot \nabla \right) \bmath{u} \right],
\end{equation}
\begin{equation}
	\label{eq:dadt_nu}
	\left( \frac{\partial A}{\partial t} \right)_{scatt} =
	- \nu_S \left( 2A^2 - A - 1 \right),
\end{equation}
where $A=T_{\perp}/T_{\parallel}$ is the ratio between the temperature components;
$\bmath{u}$ and $\bmath{B}$ are respectively the velocity and magnetic fields, with $\bmath{b = B}/B$; and 
$\nu_{S}$ is the pitch-angle scattering rate. The ions and electrons were considered 
to have identical temperature components, for simplicity. Also for simplicity, the cooling 
employed was considered  not to affect the temperature anisotropy.

The effective scattering rate $\nu_{S}$ accounts for the effect of both the Coulomb collisions and the 
non-linear particle-plasma wave interactions. 
In \citetalias{santos-lima_etal_2014} the Coulomb collisions were neglected and the scattering was attributed 
only to the action of the mirror and firehose instabilities whenever the anisotropy $A$ 
overcame the threshold values for these instabilities. Different values were considered for $\nu_S$, 
from the limit of no scattering ($\nu_S = 0$) till the extreme case in which the scattering time is very  short  or 
infinitely small
compared to the resolved timescales of the simulation ($\nu_S = \infty$).

Our purpose here is (i) to provide an evaluation of the scattering rate $\nu_S$  
due to the plasma instabilities, and (ii) to estimate the limits on the ions anisotropy $A_i$ in the ICM plasma 
by imposing the statistical equilibrium between the terms 
$\left( \partial A_i / \partial t \right)_{CGL}$ and
$\left( \partial A_i / \partial t \right)_{scatt}$.
For this aim we will follow  three steps: 
(1) obtain from the MHD turbulence simulation 
the characteristic time for the anisotropy development
$\tau_{A} = A_i \langle \partial A_i / \partial t \rangle^{-1}_{CGL}$
as a function of the ion plasma parameters $A_i$ and $\beta_{i\parallel}$;  
(2)  estimate $\nu_S (A_i, \beta_{i\parallel})$ using quasilinear 
theory and then calculate the characteristic time for the anisotropy relaxation  
$\tau_{\nu} = A_i \langle \partial A_i / \partial t \rangle^{-1}_{scatt}$;
(3) find the values of $A_i(\beta_{i\parallel})$ for which 
$\tau_{A} = \tau_{\nu}$, in order to estimate 
the maximum anisotropy level that the turbulence can sustain in the presence of 
the instabilities scattering~\footnote{Rigorously speaking, 
the maintenance of the marginal state during 
the simultaneous anisotropy driving and relaxation should also take into account the 
evolution of the local magnetic field intensity. For example, the mirror instability is set theoretically 
for $A > 1 + \beta_{\perp}^{-1}$; therefore,
to keep the plasma in the marginal state requires
$(\partial A / \partial t)_{CGL} + (\partial A / \partial t)_{scatt} \le (\partial \beta_{\perp}^{-1} / \partial t)$.}.

The AMHD turbulence simulation we employ in step (1) has $\nu_S=0$ (wich is non realistic
as it will be seen). It corresponds to the model $A1$ presented in \citetalias{santos-lima_etal_2014}. 
The value of $\nu_S$ of the MHD simulation 
is of little importance in this stage  because it should not influence the  evaluation of 
$\tau_{A}$ (at least in order of magnitude), 
though it changes 
considerably the spreading of the PDF of the plasma parameters $(A, \beta_{\parallel})$.
To confirm this, we also repeated our
analysis using an AMHD model with a physically more plausible value of $\nu_S$
($\nu_S \sim 10 \tau_{turb}^{-1}$; see below).
We consider an uniform magnetic field  in the domain; the ratio between the 
unperturbed thermal pressure and the magnetic pressure of this uniform magnetic field has the value of 
$\beta_0 = 200$, which is representative of the ICM. Super-Alfvenic and subsonic turbulence 
(with Alfvenic Mach number $M_A \equiv \langle |\bmath{u}| / v_A \rangle \approx 1.2$ and 
sonic Mach number $M_S \equiv \langle |\bmath{u}| / c_S \rangle \approx 0.6$) 
is considered  
with an  injection scale $l_{turb} = 0.4 L_0$, where $L_0$ is the computational box size. 
The employed resolution ($512^3$) allows for solving a modest inertial range covering the range 
of dimensionless wavenumbers $2.5 \lesssim k L_0 \lesssim 20$.
Further details on the numerical setup, code, and the turbulence statistics analysis can 
be find in \citetalias{santos-lima_etal_2014}.

We define the following physical dimensions for our simulations: 
$L_0 = 100 \; {\rm kpc}$ is the box size, 
$\rho_0 = 10^{-27} \; {\rm g/cm}^{3}$ is the mean density, and 
$c_{S0} = 10^8 \; {\rm cm/s}$ is the unperturbed thermal speed 
(corresponding to the gas temperature $T_0 \approx 6 \times 10^8$ K).
With this choice of units, $l_{turb} = 40 \; {\rm kpc}$, $u_{rms} \approx 7 \times 10^7 \; 
{\rm cm/s}$, and $B_0 = 3 \; \mu {\rm G}$ is the intensity of the mean magnetic field, 
corresponding to the ion thermal gyrofrequency $\Omega_{i0} \approx 3 \times 10^{-2} \; {\rm s}^{-1}$.

\section{Quasilinear evolution of the kinetic instabilities}
\label{sec:ql}

The electromagnetic waves in the plasma can interact with the particles, 
exchanging energy and momentum. This process can be described statistically 
as a diffusion of the distribution function in the velocity space.

In a collisionless plasma composed by ions (protons) and electrons, 
the electromagnetic fluctuations driven by thermal anisotropy more important for the scattering of 
the ions are generated by the firehose, mirror, and ion-cyclotron instabilities \citep{gary_1993}.
The firehose instability can be excited when $T_{i\perp} < T_{i\parallel}$, and the mirror and ion-cyclotron 
can be excited in the opposite regime $T_{i\perp} > T_{i\parallel}$.
The resulting scattering from these instabilities decreases the temperature anisotropies and consequently 
regulates the growth of the instabilities themselves.
The fastest growth modes for these instabilities occur for scales close to the 
ion Larmor radius, with growth rates which can be of the order of the ion Larmor frequency.
The electrons anisotropy is expected to be relaxed on faster time-scales (by the whistler and firehose modes; 
see \citealt{gary_1993, nishimura_etal_2002, stverak_etal_2008}).

The non-linear development of the instabilities can be investigated analytically 
using the quasilinear theory, which assumes small perturbations of the distribution functions and 
of the electromagnetic fields (compared to the zeroth order, background values).
The quasilinear theory also assumes the superposition of non-interacting plasma waves 
with random phases, which satisfy the linear dispersion relation of the plasma. 
The second order effects of these waves on the particles distribution function give rise 
to a diffusion term in the momentum space, which can be interpreted as resulting from 
effective collisions. 
In Section~\ref{sec:limitations_ql} we discuss the limitations of the quasilinear approximation 
to approach the instabilities evolution.

\citet{hellinger_etal_2013} provide the general quasilinear expressions for the evolution of the
mean velocity and thermal energy components of a general drifting bi-Maxwellian plasma composed by 
protons and electrons. Here we will use the simpler expressions derived in \citet{yoon_seough_2012} and 
\citet{seough_yoon_2012} for a bi-Maxwellian  distribution function for the ions and an isotropic distribution 
for the electrons, 
for the evolution of the temperature components due to the parallel firehose, 
mirror, and ion-cyclotron instabilities. Below we reproduce these 
expressions.

\subsection{Parallel firehose modes}
\label{sec:ql_eqs_firehose}

The linear dispersion relation for the firehose modes (modes with right-hand circular polarization) 
propagating parallel to the mean magnetic field is given by:
\begin{multline}
0 = \frac{c^2 k^2}{\omega_{pi}^{2}} - 
\frac{\omega_k}{\Omega_i} + \left( 1 - \frac{T_{i\perp}}{T_{i\parallel}} \right) \\
- \left[ \frac{T_{i\perp}}{T_{i\parallel}} \omega_k - 
\left( 1 - \frac{T_{i\perp}}{T_{i\parallel}} \right) \Omega_i \right]
	\frac{1}{k v_{i\parallel}} Z \left( \frac{\omega_k + \Omega_i}{k v_{i\parallel}} \right)
\end{multline}
where $\omega_k = \omega(k)$ is the wave complex frequency for the wave-vector
$k = k_{\parallel}$, $\omega_{pi} = \sqrt{4 \pi n_i e^2/m_i}$ and
$\Omega_i = e B_0 / m_i c$ are respectively the 
plasma frequency and Larmor frequency for the ions,
$v_{i\parallel} = \sqrt{T_{i\parallel} / m_i}$ is the parallel thermal speed of the ions,
$Z (\xi)$ is the plasma function; 
$n_i, e, B_0$, and $m_i$ are the ions density, elementary charge, background magnetic field intensity, 
and ion mass, respectively.
The terms of the order $(\omega_{k}/\omega_{pi})^2$ and $\omega_{k}/\Omega_{e}$ were neglected 
in the above dispersion relation.

The evolution equations for the ion kinetic energies (second order moments of the distribution function) 
resulting from the interaction with the 
parallel firehose modes are given by
\begin{equation}
	n_i \frac{d T_{i\perp}}{d t} = 8 \int_0^{\infty} dk \gamma_k 
	\frac{|\delta B_k|^2}{8 \pi} \left[ 
	\frac{\Re(\omega_k) \Omega_i}{k^2 v_A^2} - 1 \right],
\end{equation}
\begin{equation}
	\frac{n_i}{2} \frac{d T_{i\parallel}}{d t} = - 4 \int_{0}^{\infty} dk \gamma_k 
	\frac{|\delta B_k|^2}{8 \pi} \left[ 
	\frac{2 \Re(\omega_k) \Omega_i}{k^2 v_A^2} - 1 \right],
\end{equation}
where $v_A = B_0 / \sqrt{4 \pi n_i m_i}$ is the Alfven velocity, 
$\Re(\omega_k)$ and $\gamma_k$ are the real and imaginary parts of $\omega_k$, respectively, 
and $|\delta B_k|^2 / 8 \pi$ is the spectral energy density of the magnetic fluctuations, which 
evolves accordingly to the wave kinetic equation
\begin{equation}
	\frac{\partial |\delta B_k|^2}{\partial t} = 2 \gamma_k |\delta B_k|^2.
\end{equation}

We refer to \citet{seough_yoon_2012} for more details on the deduction of the above equations.

\subsection{Ion-cyclotron and mirror modes}
\label{sec:ql_eqs_ic}

The linear dispersion relation for the ion-cyclotron modes (with left-hand side polarization) 
propagating in an arbitrary oblique direction to the mean magnetic field is given by
\begin{multline}
0 = \frac{c^2 k^2}{\omega_{pi}^{2}} + \frac{\omega_{\bmath k}^{IC}}{\Omega_i} - 2 \frac{I_1(\lambda) \exp(-\lambda)}{\lambda} \times \\
\times \left[ \xi^{IC} Z(\zeta^{IC}) - \left( \frac{T_{i\perp}}{T_{i\parallel}} - 1\right) \frac{Z'(\zeta^{IC})}{2} \right],
\end{multline}
where $\omega_{\bmath k}^{IC} = \omega^{IC} ({\bmath k})$ is the complex wave-frequency for the two-dimensional wave-vector
${\bmath k} = (k_{\parallel}, k_{\perp})$, $\xi^{IC} = \omega_{\bmath k}^{IC}/k_{\parallel} v_{i\parallel}$,
$\zeta^{IC} = (\omega_{\bmath k}^{IC} - \Omega_i)/k_{\parallel} v_{i\parallel}$, 
$\lambda = k_{\perp}^{2} v_{i \perp}^{2}/2\Omega_i^2$, $v_{i\perp} = \sqrt{2 T_{\perp} / m_i}$,  $I_j(\lambda)$ is the 
modified Bessel function of the first kind of order $j$, and $Z'$ is the derivative of the plasma function $Z$. 

The dispersion relation of the non-propagating mirror modes is in turn given by
\begin{multline}
0 = \frac{c^2 k^2}{\omega_{pi}^{2}} + 2 \lambda \left[ I_0(\lambda) - I_1(\lambda)\right] \times \\
\times \exp(-\lambda) \left[ 1 + \frac{T_{i\perp}}{T_{i\parallel}} \frac{Z'(\xi^{M})}{2} \right],
\end{multline}
where $\xi = i \gamma_{\bmath k}^{M}/k_{\parallel} v_{i\parallel}$. Like for the firehose instability, 
the terms of the order $(\omega_{\bmath k}/\omega_{pi})^2$ and $\omega_{\bmath k}/\Omega_{e}$ were neglected 
in the dispersion relation for the ion-cyclotron and mirror modes.

The equations describing the evolution of the ion kinetic energy components are given by
\begin{multline}
	\label{eq:tperp_ic}
	n_i \frac{d T_{i\perp}}{d t} = 
        - 16 \pi \int_0^{\infty} dk_{\parallel} \int_0^{\infty} k_{\perp} dk_{\perp} \times \\
        \times \Bigg\{
        \gamma_{\bmath k}^{IC} \frac{|\delta B_{\bmath k}^{IC}|^2}{8 \pi} \left[ 
	1 + \left( \frac{\Re(\omega_{\bmath k}^{IC})}{\Omega_i} - \frac{1}{2} + \frac{\Lambda_1}{\lambda} \right)
	\frac{\Omega_i^2}{k^2v_A^2} \right] \\
        + \gamma_{\bmath k}^{M} \frac{|\delta B_{\bmath k}^{M}|^2}{8 \pi} \left[ 
	1 + \lambda \left( \Lambda_0 - \Lambda_1 \right) \frac{\Omega_i^2}{k^2v_A^2} \right] 
        \Bigg\},
\end{multline}
\begin{multline}
	\label{eq:tparallel_ic}
	\frac{n_i}{2} \frac{d T_{i\parallel}}{d t} = 
        8 \pi \int_0^{\infty} dk_{\parallel} \int_0^{\infty} k_{\perp} dk_{\perp} \times \\
        \times \Bigg\{
        \gamma_{\bmath k}^{IC} \frac{|\delta B_{\bmath k}^{IC}|^2}{8 \pi} \left[ 
	1 + 2 \left( \frac{\Re(\omega_{\bmath k}^{IC})}{\Omega_i} - \frac{1}{2} + \frac{\Lambda_1}{\lambda} \right) 
	\frac{\Omega_i^2}{k^2v_A^2} \right] \\
        + \gamma_{\bmath k}^{M} \frac{|\delta B_{\bmath k}^{M}|^2}{8 \pi} \left[ 
	1 + 2 \lambda \left( \Lambda_0 - \Lambda_1 \right) \frac{\Omega_i^2}{k^2v_A^2} \right] 
        \Bigg\},
\end{multline}
where we used the definition $\Lambda_j = I_j(\lambda) \exp(- \lambda)$. 
The kinetic wave equations for the ion-cyclotron 
and mirror modes are
\begin{equation}
	\label{eq:wave_ic}
	\frac{\partial |\delta B_{\bmath k}^{IC}|^{2}}{\partial t} = 2 \gamma_{\bmath k}^{IC} |\delta B_{\bmath k}^{IC}|^{2}.
\end{equation}
\begin{equation}
	\label{eq:wave_mirror}
	\frac{\partial |\delta B_{\bmath k}^{M}|^{2}}{\partial t} = 2 \gamma_{\bmath k}^{M} |\delta B_{\bmath k}^{M}|^{2}.
\end{equation}
The derivation of the above equations can be found in \citet{yoon_seough_2012}.

\subsection{Numerical methods}
\label{sec:ql_numerical}

The quasilinear equations for the evolution of the ions temperature components and  
of the magnetic energy modes were integrated using the LSODE 
solver from the numerical library ODEPACK \citep{hindmarsh_1983, radhakrishnan_hindmarsh_1993}.
At each iteration, the linear dispersion equation for each instability is solved numerically 
inside a discrete domain $(k_{\parallel}(i), k_{\perp}(j))$ defined by 
$k_{\parallel,\perp}(i) = (i - 0.5) * k_{\max}/N$ ($1 \le i \le N$), 
where
$0 < k_{\max} r_i < 2$, $r_i$ is the thermal ion Larmor radius, and $N=256$. 
For the firehose modes, only the unidimensional grid $k_{\parallel}(i)$ was used.
For all the calculations presented, a flat spectrum of magnetic fluctuations
$| \delta B_{\bmath k} |^2 / B_0^2 = 10^{-7}$ is imposed at the beginning of the simulation.

\section{Results}
\label{sec:results}

\begin{figure}
	\begin{tabular}{c c} 
	\input{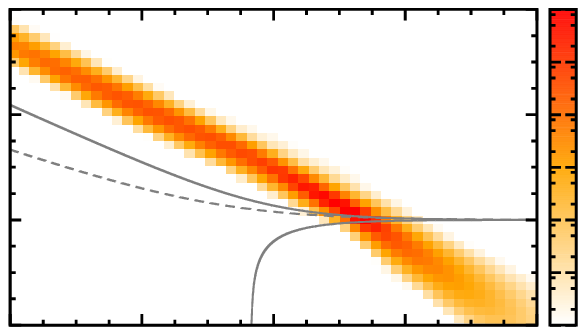} \\
	\input{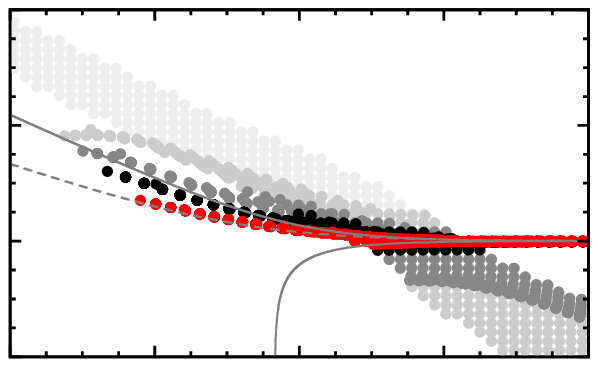} 
	\end{tabular}	
	\caption{Top: probability distribution function (PDF) for the 
		macroscopic plasma parameters $\beta_{i\parallel} = 8 \pi n_i T_{i\parallel} / B^2$ and 
		$A_i = T_{i\perp}/T_{i\parallel}$ obtained from the statistically stationary 
		state of forced turbulence of the simulation using the CGL-MHD approximation 
		by \citet[model $A2$ there]{santos-lima_etal_2014}.
		Bottom: initial values of $\beta_{i\parallel}$ and  $A_i$ from the quasilinear 
		calculations (lighter gray dots) and  values of the same parameters after the time interval 
		$\Omega_i t = 500$ (red dots). 
		The successively darker gray dots represent the system at the times $t = 10$, $20$, and $40 \; \Omega_i^{-1}$.
		The gray solid lines represent the thresholds for the  mirror 
        $A_i = 1 + 0.87\beta_{i\perp}^{-0.56}$ ($A_i>1$) 
        and parallel firehose $A_i = 1 - 0.61\beta_{i\parallel}^{-0.63}$ ($A_i<1$) instabilities; 
        the gray dashed line represents the threshold 
		for the ion cyclotron (IC) instability $A_i = 1 + 0.43\beta_{i\perp}^{-0.42}$ 
        (all the thresholds are obtained from linear theory; 
	see \citealt{seough_yoon_2012} and references therein).}
	\label{fig:map_pdf}
\end{figure}

\begin{figure}
	\begin{tabular}{c}
	\input{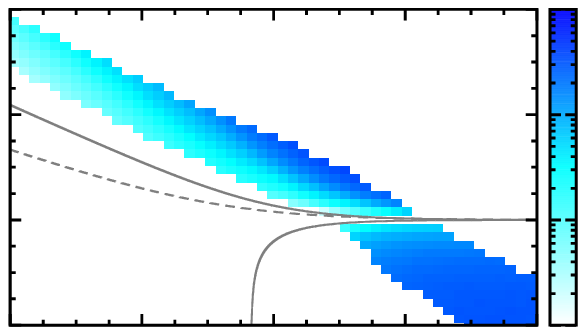} \\
	\input{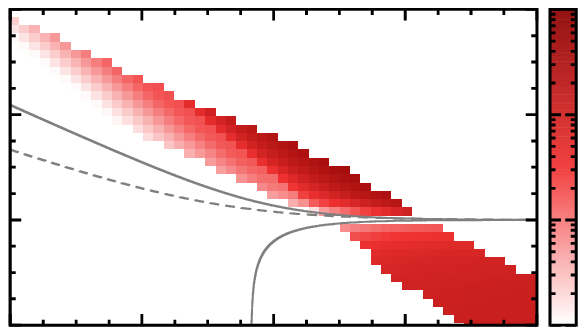} 
	\end{tabular}	
	\caption{Top: ions scattering rate averaged in time $\langle \nu_{S} \rangle$ (normalized by the 
		Larmor frequency $\Omega_i$) for each initial state 
		($\beta_{i\parallel}$, $A_i$) of the quasilinear evolution. The average in $\nu_{S}$
                only considers
                times for which $\nu_S \ge 0.6 \nu_{\max}$, where $\nu_{\max}$ is the 
		maximum scattering rate during the system evolution.
		Bottom: maximum magnetic energy density in the ion-cyclotron + mirror 
		($A_i>1$) and firehose ($A_i<1$) modes during the quasilinear evolution of 
		each initial state ($\beta_{i\parallel}$, $A_i$). The gray lines have
	the same meaning as in Figure~\ref{fig:map_pdf}.}
	\label{fig:map_nu}
\end{figure}

\begin{figure}
	\begin{tabular}{c}
	\input{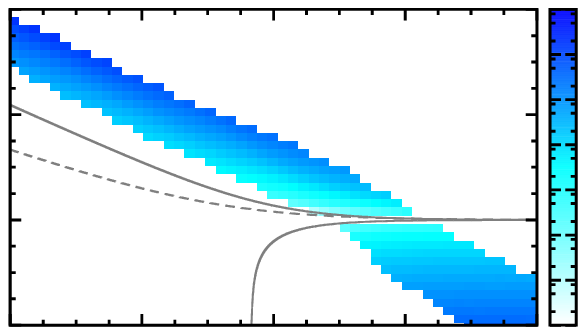} \\
	\input{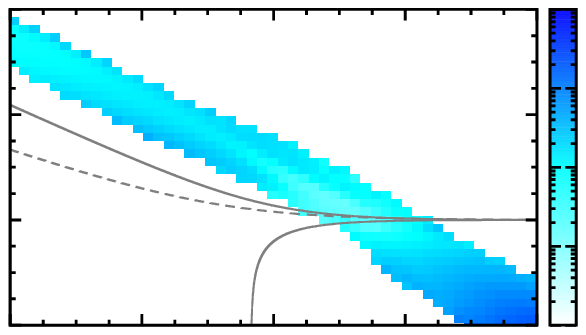} 
	\end{tabular}	
	\caption{Top: characteristic rate of anisotropy relaxation (normalized by the proton 
	Larmor frequency $\Omega_i$) due to the instabilities scattering 
	$\Gamma_{\nu} = | \left( \partial A_i / \partial t \right)_{\nu} | A^{-1}$ 
	calculated using the average quasilinear scattering rates $\langle \nu_S \rangle$ (see Eq.~\ref{eq:dadt_nu}). 
	Bottom: characteristic rate of anisotropy increase (for $A_i>1$) or decrease (for $A_i<1$) obtained 
	from the CGL-MHD turbulence simulation of Figure~\ref{fig:map_pdf} (top), 
	$\Gamma_A = | \langle \partial A_i / \partial t \rangle_{CGL} | A_i^{-1}$ normalized by the 
	proton Larmor frequency $\Omega_{i0}$ of the mean magnetic field $B_0$ (see Eq.~\ref{eq:dadt_cgl}). 
	The average was performed using only the plasma volume where the 
	anisotropy $A_i$ was increasing for $A_i>1$ and decreasing for $A_i<1$. 
    }
	\label{fig:map_rates}
\end{figure}

\begin{figure}
	\begin{tabular}{c}
	\input{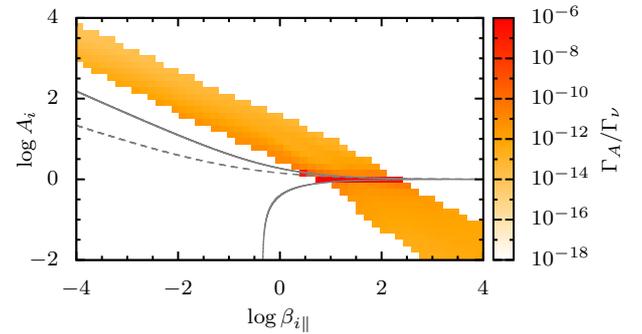} 
	\end{tabular}	
	\caption{Ratio between the characteristic rate of anisotropy change obtained from the CGL-MHD
		turbulence simulation $\Gamma_{A} = | \langle \partial A_i / \partial t \rangle_{CGL} | A_i^{-1}$ 
 	        and the characteristic rate of anisotropy relaxation calculated from quasilinear theory
		$\Gamma_{\nu} = | \left( \partial A_i / \partial t \right)_{\nu} | A_i^{-1}$
		(both presented in Figure~\ref{fig:map_rates}).}
	\label{fig:map_ratio}
\end{figure}

\begin{figure}
	\begin{tabular}{c}
	\input{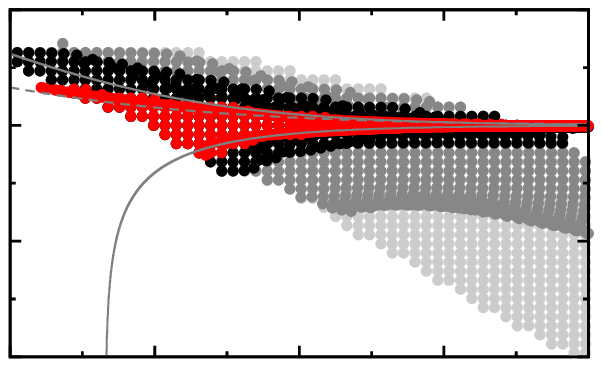} \\
	\input{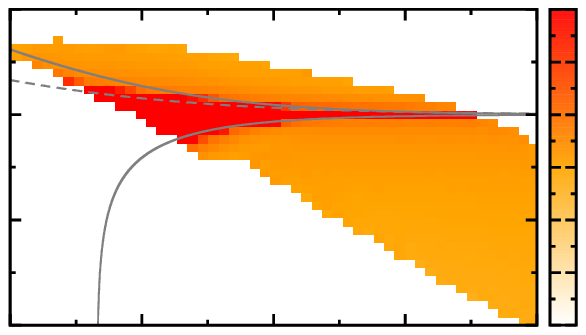}
	\end{tabular}	
	\caption{Top: same as Figure~\ref{fig:map_pdf} (bottom panel), but using 
		the initial values of ($\beta_{i\parallel}$, $A_i$) obtained from a 
		turbulence AMHD simulation which employed a non-null, constant rate $\nu_S$ in the equation of 
		anisotropy evolution (model $A3$ of \citetalias{santos-lima_etal_2014}). 
		Here the gray dots correspond to the times $t=0$, $20$, and $40 \; \Omega_i^{-1}$ (from the lighter to the darker ones).
	Bottom: same as in Figure~\ref{fig:map_ratio}, but  for the model above.}
	\label{fig:map_pdf2}
\end{figure}

Top panel of Figure~\ref{fig:map_pdf} shows the probability distribution function (PDF) of the macroscopic dimensionless 
plasma parameters $\beta_{i\parallel} = 8 \pi n_i T_{i\parallel} / B^2$ and $A_i = T_{i\perp}/T_{i\parallel}$ for 
the CGL-MHD numerical simulation of forced turbulence described in Section~\ref{sec:amhd_simul} 
(i.e., model A2 of \citetalias{santos-lima_etal_2014} with null scattering rate $\nu_S=0$).
Most of the plasma volume has the parameters ($\beta_{i\parallel}, A_i$) inside the unstable 
zones. The thresholds for the mirror 
and firehose instabilities are represented  in Figure~\ref{fig:map_pdf} 
by the continuous gray lines and  the threshold for the 
parallel ion-cyclotron (IC) instability is represented   by the dashed gray line. 
We note that   the threshold for the ion-cyclotron instability is  more constraining  
than that of  the mirror instability in the regime $\beta_{i\parallel} < 1$.

For a grid of values ($\beta_{i\parallel}, A_i$)
where the PDF of the CGL-MHD simulation is above an arbitrary cutoff of $10^{-7}$
(lighter gray dots in the bottom panel of Figure~\ref{fig:map_pdf}), 
we calculated
the quasilinear evolution of the ion temperatures $T_{i\perp}$ and $T_{i\parallel}$ 
due to the wave-particle scattering of the ions 
by the parallel firehose, mirror, and ion-cyclotron modes.
The evolved values of ($\beta_{i\parallel}$, $A_i$) after a time interval $\Omega_i t = 500$
are shown as red dots in the bottom panel of Figure~\ref{fig:map_pdf}.
For each initial condition, the plasma parameters evolved to values close to the 
marginal equilibrium state.

Top panel of 
Figure~\ref{fig:map_nu} depicts the ions scattering rates $\langle \nu_{S} \rangle$ (normalized by the ion Larmor 
frequency $\Omega_i$) resulting from the quasilinear evolution,  as a function of the initial states ($\beta_{i\parallel}, A_i$).
These scattering rates were obtained from a temporal average of the 
instantaneous scattering rates, taking 
into account only values of $\nu_S \ge 0.6 \nu_{\max}$, where $\nu_{\max}$
is the maximum scattering rate obtained during the time evolution.
The values of $\langle \nu_{S} \rangle/ \Omega_i$ are mostly in the interval $10^{-2}-10^{-1}$, but 
 inside the stable region they drop quickly to zero (this cannot be visualized in the Figure since the values 
of $\langle \nu_{S} \rangle$ fall below the color scale range at the region near $A_i \sim 1$ ).
We further notice that the values of $\langle \nu_{S} \rangle/\Omega_i$ increase with $\beta_{i\parallel}$. 

The bottom panel of Figure~\ref{fig:map_nu} shows, as a function of the initial states ($\beta_{i\parallel}, A_i$), 
the maximum magnetic energy density in the modes (ion-cyclotron, mirror, and firehose)
during the quasilinear evolution, normalized by the energy density of the 
background magnetic field $B_0$. 
For most of the initial conditions, this quantity is below  unity 
and does not break the assumption of small perturbations of the Larmor orbit. 
However, for initial conditions far from the thresholds (specially in the high-$\beta_{i\parallel}$ 
region for $A_i<1$), it achieves values of the order or larger than 1. 
For this region, the values of $\langle \nu_{S} \rangle$ shown in the top panel of Figure~\ref{fig:map_nu} 
must be taken with caution (see Section~\ref{sec:limitations_ql}). 
Nonetheless, these same initial conditions are not expected to be  accessible by the 
ICM plasma if the wave-particle scattering is taken into account during the CGL-MHD evolution
(see below).

Top panel of Figure~\ref{fig:map_rates} shows the characteristic rate of the anisotropy relaxation
caused by the scattering due to the instabilities 
$\Gamma_{\nu} = |\left( \partial A_i / \partial t \right)_{\nu}| A_i^{-1}$ 
(normalized by the ion Larmor frequency $\Omega_i$) as a function of the 
initial plasma parameters ($\beta_{i\parallel}, A_i$), according to 
Eq.~\ref{eq:dadt_nu} and $\langle \nu_S \rangle$ from
the quasilinear calculations.
The characteristic rate at which the anisotropy  changes
in the CGL-MHD turbulence simulation described above, $\Gamma_A = |\langle \partial A_i / \partial t \rangle_{CGL}| A_i^{-1}$ 
(normalized by the ion Larmor frequency $\Omega_{i0}$ 
in the uniform magnetic field $B_0$; 
see Eq.~\ref{eq:dadt_cgl}), is shown in the bottom panel of Figure~\ref{fig:map_rates}. 
The average in $\langle \partial A_i / \partial t \rangle_{CGL}$ considers only 
the plasma volume of the simulation with $(\partial A_i / \partial t)_{CGL} > 0$ when $A_i>1$
and $(\partial A_i / \partial t)_{CGL} < 0$ when $A_i<1$, in order to capture the rate 
at which the anisotropy is driven apart from the stable zone.
It is evident that $\Gamma_{\nu} \gg \Gamma_{A}$
for all the unstable region.~\footnote{However, the values of 
$\Gamma_{A}$ from the simulations are expected to increase with resolution; the average value obtained 
in the simulation presented here could be until 6 orders of magnitude below the real one (see discussion 
in Section~\ref{sec:applicability_ba}). Even taking into account this possible big difference, 
the inequality $\Gamma_{\nu} \gg \Gamma_{A}$ is still largely valid.}

It is clear that the maximum and minimum values of $A_i$
that the turbulence can sustain are limited by the temperature anisotropy relaxation rates
due to the instabilities. By comparing $\Gamma_{A}$ and $\Gamma_{\nu}$, we can find
for each value of $\beta_{i\parallel}$ the maximum/minimum values of $A_i$ ($A_i^{\pm}$)
from the balancing $\Gamma_{A} (A_i^{\pm}) = \Gamma_{\nu} (A_i^{\pm})$. Figure~\ref{fig:map_ratio} 
shows the ratio $\Gamma_{A} / \Gamma_{\nu}$ between the rates presented in Figure~\ref{fig:map_rates}.
The separation of $A_i^{\pm}$ from the mirror and firehose thresholds cannot be 
resolved for the grid in the ($\beta_{i\parallel}, A_i$)-plane used in our calculations.
However, it is evident that this separation is $\ll 1$.
It shows that the turbulence can only sustain values of the temperature anisotropy $A_i$ which are
extremely close to the instabilities thresholds.
Therefore the anisotropy levels featuring in the CGL-MHD simulation 
for the ICM turbulence are far from realistic.

We repeated all the above analysis, but now replacing the CGL-MHD simulated model used so far 
(model A2 of \citetalias{santos-lima_etal_2014} with null $\nu_S$)   for 
a simulated AMHD model in which  a non-null constant value of 
$\nu_S$ was employed (model $A3$ of \citetalias{santos-lima_etal_2014}, 
with 
$\nu_S \sim 10 u_{rms}/l_{turb}$).

Figure~\ref{fig:map_pdf2} presents  
the evolution of $(\beta_{i\parallel}, A_i)$ (top panel), 
and the ratio between the 
anisotropy change rate $\Gamma_{A} = | \langle \partial A_i / \partial t \rangle_{CGL} | A_i^{-1}$ 
and the characteristic rate of anisotropy relaxation 
$\Gamma_{\nu} = | \left( \partial A_i / \partial t \right)_{\nu} | A_i^{-1}$. 
The results of the quasilinear evolution calculation are now similar to those of  the simulated CGL-MHD turbulence model with 
the balancing between the rates $\Gamma_{A}$ and $\Gamma_{\nu}$ very close to the 
thresholds for the instabilities.

\section{Discussion}
\label{sec:discussion}

\subsection{Limitations of the CGL-MHD model to describe compressible modes}
\label{sec:limitations_cgl}

The CGL closure provides the simplest fluid model for a collisionless plasma, 
and assumes no heat flux. In particular, the linear dispersion of the 
CGL-MHD equations is known to deviate from the 
long wavelength limit of the
kinetic theory for compressible modes, 
resulting in a different threshold for the mirror instability (being over-stable compared 
to the threshold obtained from the kinetic theory).
Besides, for simplicity, we considered a CGL-MHD model with the same anisotropy in temperature 
and total thermal energy for both the ions (protons) and electrons (see discussion in \citetalias{santos-lima_etal_2014} 
and below).

Another serious limitation of the CGL-MHD model is that it does not capture the collisionless 
damping effects of the compressible modes (see, e.g. \citealt{yan_lazarian_2004}). 
Alternative higher order closures exist which can mimic
the Landau damping of the compressible modes, at least for a narrow range of wavelengths 
(see for example \citealt{snyder_etal_1997, sharma_etal_2006}).
In view of this, 
we should be cautious with regard to 
compressible modes cascade (and shocks) 
in CGL-MHD based models.

The spatial scale in which the collisionless thermal damping may be dominant in the 
collisionless
intracluster medium 
is $\sim$ 0.1-1~kpc \citep{brunetti_lazarian_2007} and therefore well below 
of the approximate inertial range 
of the turbulent models discussed here (between 5 and 40~kpc). 
Thus a potential influence of the Landau
damping in the models discussed here would be only in shock regions 
formed by the turbulence. 

On the other hand, if a considerable reduction of the parallel ion mean free path 
is assumed to occur
continuously in time in most of the plasma volume
--- via the scattering or magnetic trapping
of the ions by the plasma instabilities (see next sections),
this problem could be solved at least in part, because the large scale turbulence in the ICM 
would become effectively ``collisional''.
However, the knowledge of the spatial/temporal statistics of the parallel ions mean free path in 
the turbulent ICM is highly non-trivial, because the state of the micro-physical 
instabilities depends not only on the instantaneous properties of the flow and the macroscopic variables, but 
also on their evolution history (\citealt{melville_etal_2016}; see also the discussion in the next sections).

\subsection{Limitations of the quasilinear theory applied to initially unstable plasma configurations}
\label{sec:limitations_ql}

The quasilinear theory used here to calculate the evolution of the plasma instabilities arising from 
an initially unstable configuration 
has, of course, limitations, which are (at least in part) related to: (i) the linear approximations assumed, 
(ii) the assumption that the distribution function is bi-Maxwellian all the time, 
(iii) the neglect of non-linear interaction between waves, and specially (iv) the assumption of an homogeneous 
final state of plasma equilibrium.

Considering the limitation imposed by (i),
it should be pointed out that altough the quasilinear approximation 
is formally only applicable for very small perturbations,
the thermal ions are not sensitive to perturbations much larger than their gyro-radius, 
which are generated also by the instabilities. Thus, 
the condition $\delta B^2 / B_0^2 \ll 1$ can be slightly relaxed, 
considering the magnetic energy of the fluctuations $\propto \delta B^2$ integrated over all the spectrum.

Recently, \citet{seough_etal_2014} performed one-dimensional Particle-In-Cell (PIC) simulations of the 
ion cyclotron instability for a limited set of initial conditions 
(with a fixed anisotropy $T_{i\perp}/T_{i\parallel} = 4$ and different values of $\beta_{i\parallel}$).
They compared the evolution of the thermal energy components and of the total magnetic energy in the 
instabilities with the quasilinear predictions, 
finding good agreement for the moderate and high beta regimes ($\beta_{i\parallel} = 1$ and $10$), for which 
the linear assumption $\delta B^2 / B_0^2 \lesssim 1$ is maintained all the time. 
In the low beta regime
($\beta_{i\parallel} = 0.1$) however, the exponential growth of the instability ceased soon after the 
waves energy reached the background magnetic energy level
(at $t \approx 50 \; \Omega_{i}^{-1}$), giving place to a nearly linear growth until the saturation.
Nonetheless, the quasilinear predictions still provided a reasonable approximation 
to the PIC experiment in this case for the evolution of the thermal anisotropy.
The authors also observed that the ions distribution function  deviates from  a bi-Maxwellian 
during the early stages of the instability evolution, but this deviation vanishes at late times 
when the system achieves the stationary, saturated state (after $\sim 100\Omega_i^{-1}$).

We have also carried out comparisons of the evolution of the instabilities
obtained from two-dimensional hybrid simulations by \citet{gary_etal_2000} for a plasma with dominant perpendicular 
temperature with  quasilinear calculations
taking into account both the oblique ion cyclotron and mirror modes (see the Appendix~\ref{sec:appendix}). These results show 
good agreement (within an order of magnitude) between the scattering rates, specially for  
large values of the initial ion cyclotron growth rate. For the smallest values, the quasilinear scattering rates 
seem to overestimate the ones from the simulations. 

On the other hand, it has been verified in two-dimensional PIC and hybrid simulations the dominance of the mirror modes 
(which are oblique to the background magnetic field)
over the ion cyclotron modes for regimes of $\beta_{i\parallel} \gtrsim 1$, even when the ion cyclotron 
modes have growth rates comparable to the mirror modes \citep{kunz_etal_2014, riquelme_etal_2015}. 
These last numerical experiments focused on the situation in which the thermal energy is initially isotropic 
and one component of the external magnetic field has its intensity changed at a constant rate 
(in a shear box configuration, representing the magnetic field shearing caused by the large scale MHD turbulence motions)
driving in this way the increase of the perpendicular temperature (see Section~\ref{sec:slow_driving}).
Also employing two-dimensional PIC simulations, \citet{sironi_narayan_2015} showed that the relative role of the 
mirror and ion cyclotron instabilities is dependent of the electron to ion temperature ratio $T_e/T_i$, 
being the ion cyclotron instability dominant only when $T_e/T_i \lesssim 0.2$ for high beta plasmas (for the studied range $\beta_i \sim 5-30$). 
Even in this situation, the mirror modes can dominate after one time-scale associated to the anisotropy driving rate.
In the turbulent ICM, only a detailed modeling of the 
thermodynamical evolution of the species (taking into account 
electron-ion anomalous collisional processes) 
could provide the information on the local deviations from the 
thermal equilibrium between electrons and ions (see discussion in Section~\ref{sec:applicability_ba}). 
With respect to the global ICM properties, 
\citet{takizawa_1998, takizawa_1999} show that during the merger of sub-clusters of galaxies, 
the electrons temperature can be half of that of the ions in the post-shock ICM gas, 
in the outskirts of the cluster (where the electron-ion collision time is larger due to the lower density). 
However, these studies
considered the thermal coupling between ions and electrons mediated by Coulomb collisions only, 
and did not include any magnetic fields.

A detailed study comparing fully non-linear plasma simulations with a quasilinear approximation 
is 
still missing  for the mirror instability.
However, the stabilization mechanism of the mirror instability can be very different depending on the initial 
conditions of the temperature anisotropy. Very large anisotropies could produce modes with wavelengths 
close to the ion Larmor radius, in the case when the irreversible ions scattering is likely to 
drive the system to the marginal stability. 
However, these required levels of anisotropy can be artificially high,
like the ones generated by the CGL-MHD turbulence presented in this work.
In this scenario, the quasilinear scattering rates  
calculated here may  be considered as a  ``zeroth'' order approximation. 

For moderate values of the anisotropy beyond the threshold,
the saturated state of the mirror instability can be achieved 
by highly inhomogeneous and stable configuration of the plasma and magnetic field
\citep{kivelson_southwood_1996}, 
without breaking the magnetic momentum of the ions via anomalous scattering.
The total pressure equilibrium can be achieved by the betatron cooling of the trapped protons only 
\citep{pantellini_1998}.

Now lets us focus our attention on the plasma regime in which the parallel temperature is dominant and therefore, 
the firehose instability is present. 
\citet{seough_etal_2015} compared directly the quasilinear evolution of the parallel firehose instability 
with one-dimensional PIC simulations with  fixed initial anisotropy $T_{i\perp}/T_{i\parallel} = 0.1$ and different
values of the plasma beta parameter: $\beta_{i\parallel} = 2.5$, $5$, and $10$. 
Similar to the ion cyclotron study \citep{seough_etal_2014}, 
the quasilinear predictions provide a better agreement for the highest values of $\beta_{i\parallel}$.
However,
after a short initial phase of exponential growth where the quasilinear calculations 
are almost identical to the simulations, 
the saturation values of the magnetic energy modes predicted by the quasilinear calculations are 
found to be larger than the values obtained from the plasma simulations. 
For the lowest value of $\beta_{i\parallel}$ tested  ($\beta_{i\parallel}=2.5$), 
the agreement is the poorest 
and the final saturated value of the anisotropy is far from the threshold of the firehose instability. 
They also observed that the deviation from
the initial bi-Maxwellian velocity distribution is larger for 
smaller $\beta_{i\parallel}$. 
The authors suggest that the existence of strong wave-wave interactions 
could be responsible for the deviation from the quasilinear calculations.

The quasilinear calculations presented in this study only consider the evolution of the plasma 
instabilities from a set of initially unstable plasma configurations taken from the statistics 
of numerical simulations of CGL-MHD turbulence that did not consider self-consistently 
the feedback of the small-scale (subgrid) plasma instabilities. 
If our quasilinear rates of anisotropy relaxation due to the 
ions scattering are valid at least 
in order of magnitude, the 
straightforward conclusion 
that one can draw is that there is an obvious 
physical inconsistency in neglecting the micro-instabilities effects 
on the evolution of the temperature anisotropy in anisotropic MHD (AMHD) simulations of turbulence, 
at least for the observed conditions of the ICM. 
Even for an AMHD model with an imposed 
anisotropy relaxation rate of $\nu_{\rm eff} \approx 10 \tau_{turb}^{-1}$ 
(where $\tau_{turb}$ is the turbulence turn-over time)
is uniform over all the firehose and mirror unstable volume, 
the levels of temperature anisotropy achieved would generate micro-instabilities so strong in a real plasma
that they would bring the anisotropy to the (near) marginal state almost ``instantaneously'' 
($\sim$ ion kinetic time-scales).

\subsection{Mirror and firehose development under slow temperature anisotropy driving}
\label{sec:slow_driving}

Recent kinetic simulations 
have shed some light on the saturation mechanism of the mirror and firehose instabilities in the 
context of ``slow'' anisotropy driving, as expected by the ICM turbulence.
\citet{kunz_etal_2014} examines the development of the mirror and firehose instabilities 
in the situation where the anisotropy is continuously driven at a nearly linear rate 
$S \sim A^{-1} | \left( \partial A / \partial t \right)_{CGL} | \ll \Omega_i$
(see Eq.~\ref{eq:dadt_cgl})
by the large scale shear of the background magnetic field.
\citet{riquelme_etal_2015} did a similar study for the mirror instability only.
Both studies focus on the regime of $\beta \approx 200$, characteristic of ICM conditions.
\citet{melville_etal_2016} extends these studies to higher values of $\beta$ (relevant for the problem of magnetic field 
amplification in the ICM), and also analyses the decaying/evolution of the instabilities when the anisotropy driving ceases or is 
reverted.\footnote{The local shear rate $S \sim \delta v_{\perp} / l$ produced by turbulence in the scale $l$ is expected 
to be coherent during the cascading time of these scales $\sim l / \delta v_{\perp} \sim S^{-1}$.}

These studies clearly show that the temperature anisotropy is tightly limited by the
firehose and mirror marginal stability thresholds in the asymptotic limit $S \ll \Omega_i$. 
For the firehose instability, in the regime $(S \beta / \Omega_i) \ll 1$ (relevant for the ICM parameters), 
the anomalous scattering is set by the macroscopic anisotropy generation rate $S$ after a time delay $\delta t \ll S^{-1}$,
while in the regime when $(S \beta / \Omega_i) \gtrsim 1$ (relevant for the early scenario of magnetic field amplification in the ICM), 
the time interval $\delta t$ for the development of the magnetic fluctuations able to scatter the ions at a rate which equilibrates the anisotropy generation 
is $\delta t\gtrsim S^{-1}$ \citep{kunz_etal_2014, melville_etal_2016}. In both cases, the firehose fluctuations decay exponentially at a rate $\sim \Omega_i / \beta$
after the shutdown of the anisotropy driving \citep{melville_etal_2016}. 
In contrast, for mirror instability the magnetic fluctuations keep increasing continuously during all the shear time $S^{-1}$, 
with the maintenance of the marginal stability condition  
due to the increasing fraction of ions trapped in regions where the increase of the magnetic field is compensated 
by the magnetic fluctuations (the trapped particles do not feel the increase of the mean magnetic field
and are not subject to betatron acceleration; see
\citealt{kunz_etal_2014, riquelme_etal_2015, rincon_etal_2015}). 
These magnetic structures have $\delta B_{\parallel} \gg \delta B_{\perp}$ and are elongated in the 
direction of the local mean magnetic field. 
In the situation when the anisotropy driving is removed at $St = 1$, 
the mirror fluctuations decay at a rate $\sim \Omega_i/\beta$, slower than exponential \citep{melville_etal_2016}.

\citet{melville_etal_2016} also analysed the situation when the direction of the anisotropy driving is 
reversed after the time $S^{-1}$. 
The firehose development on the top of the reminiscent mirror modes 
proceeds very similar to its development from the homogeneous and isotropic initial condition.
In the case when the driving of excess of parallel pressure is inverted to an excess 
of perpendicular pressure, the plasma only develops enough anisotropy to trigger mirror modes after a substantial 
decaying of the firehose modes. 

In the next section we further discuss our results in the light of those above,  putting our work in a broader context.

\subsection{Applicability of the bounded anisotropy model to turbulence simulations of the ICM}
\label{sec:applicability_ba}

The physical fields evolved in our AMHD simulations of the ICM are in fact {\it mean fields}, 
in the sense that they represent macroscopic averages in space and time,
over scales much larger than those related to the firehose and mirror modes expected 
to develop there.
This macroscopic description therefore filters the ``microscale'' magnetic fluctuations 
which can achieve intensities comparable to the macroscopic magnetic field (for example, 
the firehose modes in the ``ultra-high'' beta 
regime~\footnote{The ``moderate'' and ``ultra-high'' $\beta$ regimes (respectively $\beta \ll \Omega_i/S$ and
$\beta \gtrsim \Omega_i/S$ are defined in \citet{melville_etal_2016}, where the estimatives for the
critical $\beta$ in the ICM turbulence are provided: $\beta_c \sim 10^{7-9}$, corresponding to magnetic field
intensities $\sim 10^{-9}-10^{-8}$~G. \label{foot:beta_regimes}} 
described in \citealt{melville_etal_2016}).

The most obvious complication of this description is related to the evolution of the 
macroscopic pressure components
relative to the direction of the macroscopic magnetic field. 
For example, the development 
of a microscale transverse magnetic field component 
does not change the direction (or intensity) of the mean magnetic field.
But it 
modifies the direction of the magnetic field in the small scales 
and at least part of the parallel pressure (with respect the microscopic local magnetic field)
should contribute to increase the macroscopic perpendicular pressure.
Also the changes in the magnetic field intensity due to the microscopic components
should produce changes in the macroscopic pressure anisotropy. 
In other words, the macroscopically seen thermal anisotropy {\it is} influenced
by the development of microscale magnetic field fluctuations, 
even assuming the perfect conservation of the particles magnetic moment and excluding any kinetic effect.
Summarizing, in the presence of micro-instabilities, 
the CGL closure for the mean, large scale fields is at least incomplete, 
as the microscopic effects eventually modify macroscopic thermal anisotropy evolution. 
Therefore, the inclusion of a ``subgrid'' model for the evolution of anisotropy 
in the AMHD description of the ICM turbulence is needed even in the absence of any irreversible 
scattering of the particles. 

Another complication is that anisotropy generation rate by turbulence 
increases inversely to the scale of the motions:
$\Gamma_{A} = A^{-1} | (\partial A / \partial t)_{CGL} | \sim d \ln B / dt \sim l^{-2/3}$ or $l_{\perp}^{-1/3}$ 
(considering the fast or Alfvenic/slow scaling for the velocity gradients; see 
\citealt{yan_lazarian_2011}). This means that the statistics of the anisotropy driving rate is 
strongly dependent on the inertial range of the simulation, and therefore, on the numerical
resolution.
In this way, considering the dominant scale for the anisotropy generation rate as the lowest scale
of the inertial range of our simulations ($\sim 10^{22}$~cm), and using the power law corresponding 
to the Alfvenic/slow velocity gradients to extend it to the ions kinetic scales $\sim 10^{5}$~km, 
the average value of $\Gamma_A$ from our simulations could increase by a factor 
of at most $\sim 10^{4}$.

To modify the CGL closure by imposing ``hard wall'' limits on the pressure anisotropy \citep{sharma_etal_2006}
in the AMHD description of the ICM is equivalent to assume that the relaxation of the macroscopic anisotropy 
to the instabilities threshold happens in a timescale negligible compared to the macroscopic time scales. 
This assumption is well justified for both firehose and mirror instabilities, whenever the 
rate of anisotropy generation $S$ is much smaller than the ion Larmor frequency $\Omega_i$ 
(see discussions in Sections~\ref{sec:limitations_ql} and~\ref{sec:slow_driving}), independent on the development of pitch angle scattering of the ions.
However, it also assumes that the free-energy released by the instabilities, 
which (at least in part) would be stored in microscopic magnetic fluctuations, 
is directly converted into internal energy irreversibly.
Firstly, in the case where the instabilities scatter the ions almost ``instantaneously'' during the anisotropy 
driving period, 
not necessarily all the free energy of the instability is transfered 
to the ions (or is equally distributed between ions and electrons), 
as some part of the electromagnetic energy cascades to the scales below the ion Larmor radius 
(see \citealt{kunz_etal_2014}) and ends up 
being transfered to the electrons. 
In any case, in the absence of a detailed description of the micro-turbulence cascading and of the full
thermodynamics including ion-electron collision rates, 
emission/cooling processes for each specie, electrons acceleration etc, 
it is not meaningful to pursue such detailed energy distribution in AMHD simulations of the ICM
(in the \citetalias{santos-lima_etal_2014} simulations, thermal equipartition is assumed between the ion and electrons).
Secondly, ``removing'' instantaneously the energy from the microscale magnetic fluctuations causes the magnetic energy 
pressure to be underestimated. However, 
the largest values of the relative magnetic energy of the fluctuations $\delta B^2/B^2$ 
are of the order of unity \citep{kunz_etal_2014, melville_etal_2016}, and as the 
thermal $\beta$ values relevant for the ICM are high, the magnetic field pressure is secondary 
(compared to the thermal pressure) and also dynamically unimportant in the large scales (specially in the dynamo
context).
But it could also affect the detailed energy distribution between the species  
if radiative emission would to be taken into account. 
After the anisotropy driving ceases, the microscopic magnetic fluctuations decay at a rate regulated 
by the scattering of the ions \citep{melville_etal_2016}. The magnetic energy of the fluctuations gradually released is not converted 
again in free energy of the thermal anisotropy thanks to the irreversible scattering of the ions.
As discussed in the previous sections, the mirror and firehose magnetic fluctuations decay in a time scale 
relatively short after ceased the anisotropy driving, for moderate values of $\beta$. In the 
``ultra-high'' $\beta$ regime, however, these magnetic fluctuations persist 
during dynamical time-scales.
This means that the bounded anisotropy model also cannot describe correctly the entropy evolution of the plasma.

To what extent could the ion scattering rate (and consequently the ions parallel mean free-path) be derived from 
the AMHD simulations of the turbulent ICM? Let us forget for a moment the complexity arising from the fact that the 
statistics (spatial and temporal) of the turbulent shearing/compression rates may depend on the 
micro-instabilities state (and on the resolution of the simulation, as discussed above),
and assume that the statistics of the shearing/compression is known.
As discussed before, for values of beta representative of the ICM, 
the firehose fluctuations instantaneously spark ions scattering at a rate needed to keep the macroscopic anisotropy
at the marginal threshold level, making it possible 
to derive the statistics of the scattering.
For the mirror modes (and also for the firehose modes in the regime of ``ultra-high'' beta), 
however, the 
determination of the
scattering rate depends on the knowledge of the microscopic magnetic 
fluctuations level that develops during the macroscopic time-scales of the shear/compressions. 
It means that the macroscopic fields of the plasma cannot determine the local scattering rate at a given time.

Now we consider again the influences of the ions scattering rate on the AMHD turbulence evolution itself.
In the absence of a significant decrease of the parallel mean free path of the ions,
a strong collisionless damping of the compressible modes propagating parallel to the local field can be expected. 
However, the shear velocities of the Alfven modes, transverse to the local magnetic field, are not expected to 
be affected by the ions parallel mean-free-path. That is, an MHD-like Alfvenic cascade is expected to develop 
independent of the ions parallel viscosity (see \citealt{schekochihin_etal_2005}). The linear Alfven modes are expected 
to be affected only by the  
shear viscosity component. Both the Braginskii shear viscosity $\sim r_i^2 \nu_{ii}$ (where $r_i$ is the thermal ion Larmor radius 
and $\nu_{ii}$ the ion-ion collision rate)
and the Landau
damping cannot set a viscous scale for the Alfvenic strong cascade above the ion Larmor radius
in the ICM regime of high beta plasma, subsonic turbulence (see a detailed discussion on this subject in \citealt{borovsky_gary_2009}).
On the other hand, the compressible cascade will be damped already in the much larger scales. 
This damping is of kinetic origin, and its physics cannot be captured in AMHD simulations (see Section~\ref{sec:limitations_cgl}).

Conjecturing that the coupling between the compressible and Alfvenic modes in the anisotropic MHD 
is similar to that in isotropic pressure MHD \citep{cho_lazarian_2003, kowal_lazarian_2010}, the Alfvenic cascade 
must be energetically more important 
and little affected by the compressible cascade in the ICM.
Therefore, the macroscopic turbulence statistics of the ICM should be well represented by the bounded anisotropy AMHD 
simulations
if the precise thermodynamic description is not important.

However, in the absence of significant anomalous scattering of the ions (as it is expected in the
``ultra-high'' beta regime; see \citealt{melville_etal_2016}), 
the micro-instabilities mechanism that will keep
the pressure anisotropy limited is the suppression of the streching rate of the 
magnetic field 
\citep{mogavero_schekochihin_2014}. 
This means that the
velocity and magnetic fields from the  
microinstabilities can affect the global stretching rate of the
magnetic field, and therefore, 
the small scale dynamo evolution. 
In this scenario, the mirror instabilities, for example, 
could slow down considerably
the turbulent amplification of the large scale magnetic field --- but the situation is 
more complex
because the persistent firehose fluctuations can suppress the development of the mirror 
modes \citep{melville_etal_2016}.
Such contribution from the microscales to the induction equation 
cannot be included easily in fluid simulations 
and hence, the results of the magnetic field amplification obtained from both in isotropic pressure MHD or bounded anisotropy AMHD 
previous simulations
(e.g. \citetalias{santos-lima_etal_2014}) should be taken with caution, at least in regimes of very high beta 
(see more details on this subject in \citealt{mogavero_schekochihin_2014, melville_etal_2016}).

It is worth to emphasize that several aspects of the ICM thermodynamics --- like entropy 
generation \citep{lyutikov_2007}, ion heat conduction \citep{kunz_etal_2011}, physics of the compressible modes 
(see Section~\ref{sec:limitations_cgl}), etc. --- 
which can have crucial importance for the ICM structuring and dynamics 
cannot be self-consistently approached by the boundary anisotropy AMHD model without a detailed 
modeling of the micro-instabilities evolution \citep{melville_etal_2016}.

\subsection{Implications to particle acceleration in the ICM turbulence}
\label{sec:implications_cr}

As discussed in Section~\ref{sec:applicability_ba}, a modification of the CGL-MHD equations is required
to account for the limits on the temperature anisotropy 
imposed by the thresholds of the instabilities firehose, mirror, 
and possibly ion cyclotron (in regions of low $\beta$) 
in studies of the turbulent ICM.
In \citetalias{santos-lima_etal_2014}, such limits (mirror and firehose) were shown to bring 
the turbulence statistics to be similar to the collisional MHD counterpart, and the turbulent dynamo 
was also found to amplify
the magnetic fields at rates compatible with those of the collisional MHD
(neglecting the effects of the suppresion of the streching rate of the magnetic field by 
the microinstabilities, see \S\ref{sec:applicability_ba}).

If the scattering and trapping of the ions by the instabilities makes 
the effective collisional scale of the thermal particles much smaller 
than the Coulomb ion-ion parallel collision scale over a significant fraction of the plasma,
it could justify a drastic reduction of the collisionless damping of the compressible 
modes by the thermal plasma. Invoking such a picture, \citet{brunetti_lazarian_2011} showed that the compressible modes 
can channel energy to re-accelerate efficiently relativistic particles in the ICM. In the absence 
of such anomalous scattering, only ten percent of the energy in the fast 
modes is available to accelerate the particles \citep{petrosian_etal_2006, brunetti_lazarian_2007}.

However, as discussed in Section~\ref{sec:slow_driving}, a knowledge of the magnetic fluctuation 
level of the micro-scale mirror modes 
(and also of the firehose modes in the ``ultra-high'' beta regime; 
see Footnote~\ref{foot:beta_regimes}) is required  
in order to make an estimative of the trapped fraction and scattering rate of the ions. 
In fact, as the evolution of the mirror modes occurs during the macroscopic 
time scales, the ions parallel mean free path will reduce gradually in time in the spatial location 
where the turbulence drives the anisotropy generation. 
Localized (in space and time) reduction of the ions mean free path in the turbulent plasma 
should be expected, but 
a detailed statistics (spatial/temporal distribution) 
in connection with the turbulence cascade is necessary to understand its impact on the damping 
of the compressible modes. This question deserves further extensive investigation.

\section{Summary and conclusions}
\label{sec:summary}

Previous numerical simulations of forced turbulence representing the intracluster medium regime 
showed that the turbulence can produce very high levels of anisotropy in the temperature when 
the plasma instabilities feedback is neglected (\citetalias{santos-lima_etal_2014}). 
This anisotropy 
has an important impact on the turbulence statistics, producing significant modifications when 
compared to one-temperature collisional MHD turbulence (see also \citealt{nakwacki_etal_2016} for a study 
on the impact of the pressure anisotropy on the Faraday rotation maps of the ICM). 
It also prevents the turbulent amplification of the magnetic fields, which 
is believed to be responsible for sustaining the observed intensities and coherence lengths of 
the magnetic fields present in the ICM
\citep{kotarba_etal_2011, egan_etal_2016}.

Using a grid of different initial conditions ($\beta_{i\parallel}, A_i$) taken from 
a distribution produced by a CGL-MHD simulation,
we calculated the 
non-linear 
evolution of the ions temperature components due to the 
pitch-angle scattering caused by the plasma mirror, ion-cyclotron, 
and parallel firehose instabilities using the quasilinear theory where we assumed an isotropic distribution for the  electrons temperature.
The quasilinear evolution brings the values of ($\beta_{i\parallel}, A_i$) close to the 
limits of marginal stability after a few hundred ion Larmor periods.

In counterpart, we computed the average rate at which the simulated CGL-MHD turbulence 
pushes the temperature anisotropy in the direction of unstable values. We showed that  this 
rate is 
several orders of magnitude smaller
than the rate at  which the 
pitch-angle scattering by the instabilities drives the temperature anisotropy towards the stable values, 
even when starting with small deviations from 
the instabilities thresholds.
The quasilinear evolution of the ions temperature anisotropy used here was also compared 
to that obtained from two-dimensional hybrid simulations for a small set of unstable initial conditions 
in which the plasma develops ion-cyclotron modes. 
This comparison shows good accordance (within an order of magnitude) for the rate 
of pitch-angle scattering (see Appendix~\ref{sec:appendix}).

Our quasilinear analysis 
demonstrates clearly that in the turbulent ICM, the fast scattering of the ions by the 
instabilities rules out the presence of
temperature anisotropies levels exceeding substantially the thresholds for the instabilities.
When the anisotropy level is very close to the threshold of the instabilities, 
the slow 
instability growth favors adiabatic evolution to the saturation, and then the quasilinear
scattering rates may become less representative (see Section~\ref{sec:limitations_ql} and~\ref{sec:slow_driving}). 
The last observation is particularly relevant in the case of the mirror instability.

Additionally, the development 
of the instabilities should take into account the continuous anisotropy driving 
during the macroscopic turbulence time scales. 
Such problem was recently 
addressed in the studies by \citet{kunz_etal_2014}, \citet{riquelme_etal_2015}, and \citet{melville_etal_2016}, with the conclusion 
that the instabilities induced by the ICM turbulence indeed keep the thermal anisotropy bounded by 
the instabilities thresholds during all the relevant macroscopic time-scales. 
The scattering rate 
of the ions is set ``instantaneously'' (for the macroscopic time-scales) by the firehose instability, 
and it is responsible for the anisotropy relaxation
which keeps the anisotropy limited. But this is not the case regarding the mirror instability,
for which the ions scattering increases gradually during the time-scale of the anisotropy 
generation by the turbulence shear/compression. In this case, the anisotropy is limited to the thresholds 
by processes essentially adiabatic (at least initially).

In conclusion, all these results and considerations justify the modification of the CGL-MHD equations 
for including bounds of the anisotropy at the instabilities thresholds, which is appropriate for the description of the 
ICM turbulence.
As the anisotropy relaxation rate derived from these bounds 
does not necessarily reflect the instantaneous scattering rate of the ions by the instabilities 
(at least for the mirror modes), such bounded anisotropy models cannot represent 
properly the thermodynamical evolution of the gas
nor the damping of the compressible modes, which dependend on the parallel ions mean-free-path.
Even considering these limitations (see Section~\ref{sec:applicability_ba}), the bounded anisotropy model should represent well the Alfvenic cascade
of the turbulence, assuming that thermodynamical details play no major role in the turbulence dynamics.
It has been shown in the earlier study of \citetalias{santos-lima_etal_2014} that if the temperature 
anisotropy is bounded to stable values (considering only the mirror and fireshose instabilities), 
the turbulence statistics and the magnetic field amplification due 
to the small-scale dynamo have results undistinguishable from the one-temperature MHD description
largely used in studies of the intracluster medium. 
We note however, that this last study did not take into 
account the potential effects of the microscale instabilities on the stretching rate of the magnetic field 
during the turbulent amplification which can be important at least in very high beta regimes (Section~\ref{sec:applicability_ba}).

Finally it should be emphasized that any phenomenum depending on the ion thermal parallel mean-free-path 
which can affect the intracluster medium structuring (e.g., via heat conduction, thermal instabilities, cooling),
and also cosmic-ray acceleration (see Section~\ref{sec:implications_cr}) deserve still further investigation considering the complex 
interplay between the macroscopic turbulence and the detailed evolution of the microscopic instabilities.

\section*{Acknowledgements}
\addcontentsline{toc}{section}{Acknowledgements}

RSL acknowledges partial support from a grant of the Brazilian Agency FAPESP (2013/15115-8).
RSL and HY acknowledge the support by NSFC11473006 for RSL's first visit to HY, during which part of this work was developed. 
EMGDP acknowledges partial support from FAPESP (2013/10559-5) and CNPq (306598/2009-4) grants. 
AL acknowledges the NSF grant AST 1212096 and Center
for Magnetic Self Organization (CMSO) as well as a distinguished visitor PVE/CAPES 
appointment at the Physics Graduate Program of the Federal University of Rio Grande
do Norte and thanks the INCT INEspa\c{c}o. 
RSL also thanks the kind hospitality of the Astronomy department of the University of Wisconsin/Madison
where part of this work was developed. RSL and EMGDP are indebted to  
D. Falceta-Gon\c{c}alves for useful discussions on plasma physics, and M. Kunz for the insightful 
discussions on the plasma instabilities evolution.
The authors are indebted to an anonymous referee 
for pointing out some misconceptions
in the original manuscript and for 
providing important suggestions to improve this work.







\appendix
\section{Quasilinear evolution of the ion-cyclotron instability compared to plasma simulation}
\label{sec:appendix}

\begin{figure}
	\begin{tabular}{c}
	\input{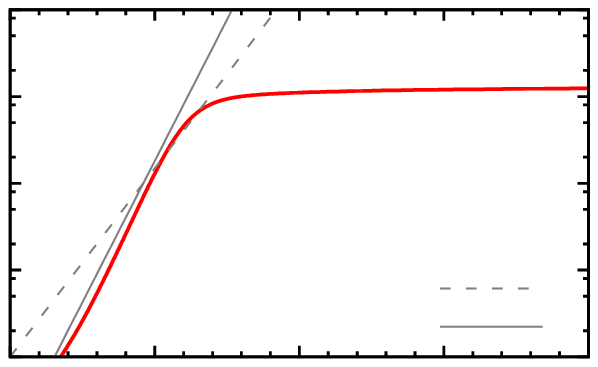} \\ 
	\input{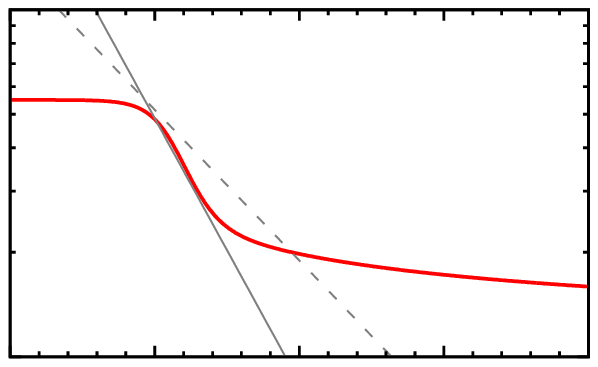} 
	\end{tabular}	
	\caption{Temporal evolution of the magnetic energy density perturbations (top) and temperature anisotropy (bottom).
	The initial conditions are $\beta_{i\parallel} = 0.05$ and $T_{i\perp}/T_{i\parallel} = 6.5$. The dashed gray
	lines represent the fitting presented in \citet{gary_etal_2000}	
	for the 2D hybrid simulations with similar initial conditions.
The continuous gray lines represent the curves which 
are the best fitting to the present quasilinear results (see text). 
}
	\label{fig:appendix_evol}
\end{figure}

\begin{figure}
	\begin{tabular}{c}
	\input{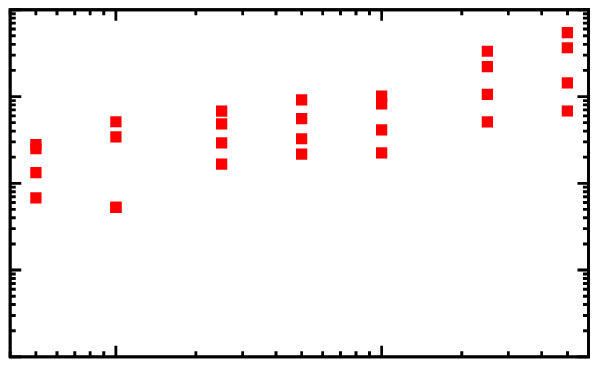} \\
	\input{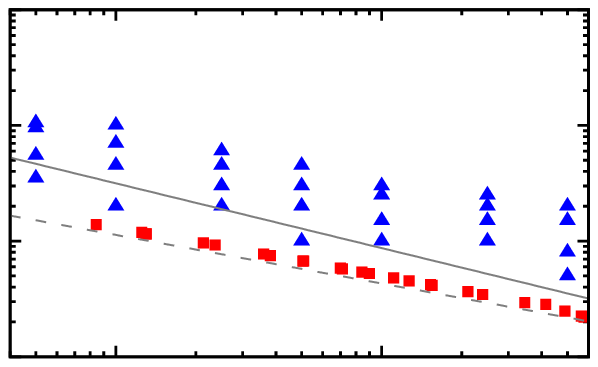} \\
	\input{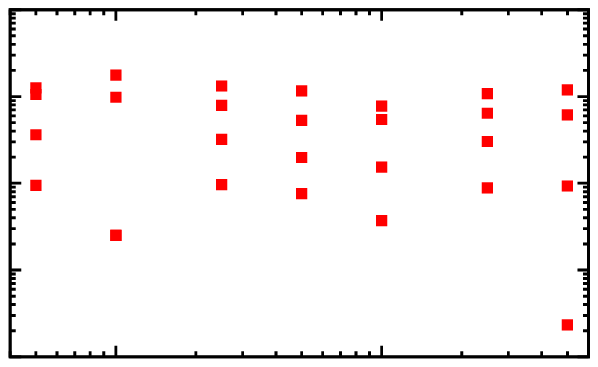} 
	\end{tabular}	
	\caption{Results from the quasilinear calculations using a set of different initial conditions
	($\beta_{i\parallel}$, $T_{i\perp}/T_{i\parallel}$) similar to those employed in the 2D hybrid simulations 
	by \citet{gary_etal_2000}.
	Top panel: maximum magnetic energy density in the ion-cyclotron modes during 
the evolution. Middle panel: initial (blue triangles) and final (red squares) temperature anisotropy;
the gray lines represent the thresholds (see Figure~\ref{fig:map_pdf}) for the ion-cylotron 
	(dashed) and mirror (continuous) instabilities.
Bottom panel: average scattering rate of the ions normalized by the 
ion Larmor frequency.}
	\label{fig:appendix_max_energy}
\end{figure}

\begin{figure}
	\begin{tabular}{c}
	\input{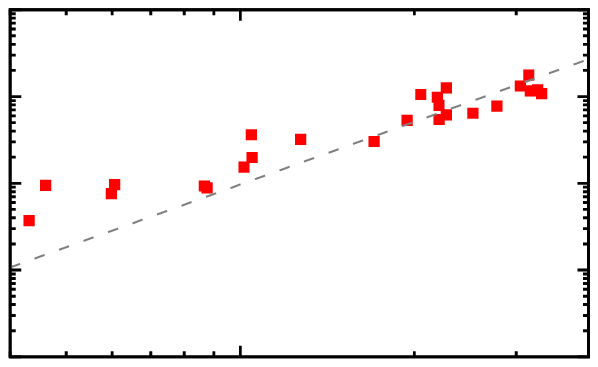}
	\end{tabular}	
	\caption{Average ions scattering rate $\langle \nu_{S}^* \rangle$ during the system evolution versus 
		the maximum growth rate $\gamma^{IC}_{\max}$ of the unstable ion-cyclotron modes.
	The dashed line represents the fitted curve in \citet{gary_etal_2000}.}
	\label{fig:appendix_nu}
\end{figure}

We present here a comparison between the evolution of the temperature anisotropy
obtained from  two-dimensional hybrid simulations presented in 
\citet[G+00 hereafter]{gary_etal_2000}, and our quasilinear 
approach described in Section~\ref{sec:ql}.

We have evolved  the ions temperature components and the
magnetic field waves energy, using  equations~\ref{eq:tperp_ic} to~\ref{eq:wave_mirror},
for a set of initial values of $\beta_{i\parallel}$ and $A_i = T_{i\perp}/T_{i\parallel}$
similar to those employed in \citetalias{gary_etal_2000}.
We considered an initial flat spectrum for the 
magnetic waves with $|\delta B_{\bmath k}|^2 / B_0^2 = 10^{-7}$.

Figure~\ref{fig:appendix_evol} shows the evolution of the energy density of the magnetic fluctuations 
$W_B = \int d{\bmath k} |\delta B_{\bmath k}|^2 / B_0^{2}$ (top panel), and the evolution 
of the ions temperature anisotropy (bottom panel) for one particular initial 
condition: $\beta_{i\parallel} = 0.05$ and $A_i=6.5$. This evolution is qualitatively similar to  
that obtained from the 2D hybrid simulation of 
\citetalias{gary_etal_2000} (see their Figure 1).
After a short time interval, $W_B$ increases exponentially. The particles scattering due to these 
magnetic fluctuations modifies the ions thermal velocity distribution and leads to a fast decrease in the anisotropy  at the same time that   the magnetic fluctuations grow very fast
(bottom panel of Figure~\ref{fig:appendix_evol}). The anisotropy 
in temperature decreases exponentially during this time interval. After this phase, it
continues to decrease but  at a slower rate.

Figure~\ref{fig:appendix_max_energy} shows for the complete set of 
the quasilinear calculations the maximum values of the energy density of the magnetic fluctuations (top panel), the initial (blue triangles) and final (red squares) values 
of the temperature anisotropy (middle panel), and the average scattering rate $\langle \nu_{S}^* \rangle$ for each simulation (bottom panel), as functions of the initial value of $\beta_{i\parallel}$. 
We note that the scattering rate $\nu_S^*$ at each time step was not calculated using Eq.~\ref{eq:dadt_nu}. Instead, in order to make a more straightforward comparison with \citetalias{gary_etal_2000}, we used:
\begin{equation}
	\label{eq:nu_gary}
	\nu_S^* \equiv - 
	\frac{1}{A_i - 1} 
	\left( \frac{\partial A_i}{\partial t} \right)_{scatt},
\end{equation}
and the time-average accounted only for values 
of $\nu_S^* \ge 0.6 \nu_{\max}^*$, where $\nu_{\max}^*$ is the maximum instantaneous scattering rate
during the time-evolution.

The results from Figure~\ref{fig:appendix_max_energy} are similar to those presented in 
Figure 2 in \citetalias{gary_etal_2000}, 
with differences no larger than one order of magnitude.

Figure~\ref{fig:appendix_nu} shows the relation between the average scattering rate $\langle \nu_{S}^{*} \rangle$ and the 
(initial) maximum growth rate of the unstable ion-cyclotron modes $\gamma^{IC}_{\max}$, for the same
set of simulations. For comparison, it is also shown the (dashed line) fitting for this relation obtained by 
\citetalias{gary_etal_2000}.
The quasilinear scattering rates agree at least in order of magnitude with the
totally non-linear calculation, and the similarity is closer for the highest initial values 
of the ion-cyclotron growth rate.


\bsp	
\label{lastpage}
\end{document}